# Effect of Magnetism on Lattice Vibrations: Mössbauer Spectroscopic Evidence

Stanisław M. Dubiel[*,1,2]

[1]Faculty of Physics and Applied Computer Science, [2]Academic Center for Materials and Nanotechnology, [1,2]AGH University of Science and Technology, PL-30-059 Kraków, Poland

**Abstract**

Based on the available literature reviewed are results pertinent to the effect of magnetism on lattice dynamics obtained by Mössbauer spectroscopy at $^{57}$Fe and $^{119}$Sn probe atoms. Presented and discussed are sigma-phase Fe-Cr, Fe-V, Fe-Cr-Ni and $\lambda$-phase (C14 Laves) NbFe$_2$ compounds that belong to the Frank-Kasper family of phases, metallic chromium and iron mono-arsenide, FeAs. The common feature of all these metallic systems is itinerant character of their magnetism. Two spectral parameters relevant to the lattice dynamics viz. the center shift, *CS*, and/or the recoil-free fraction, *f*, are considered as figures of merit. Their temperature dependences show anomalies at magnetic ordering temperatures giving evidence that that vibrations in the magnetic phase are different than those in the paramagnetic phase. The lower the magnetic ordering temperature, the stronger the difference.

**Key words:** Magnetism, Lattice Dynamics, Mössbauer Spectroscopy, Frank-Kasper Phases, Chromium, Debye Temperature

[*]Stanislaw.Dubiel@fis.agh.edu.pl

# 1. Introduction

The role of magnetism in the lattice dynamics is routinely regarded as unimportant. Such thinking follows from calculations of the standard theory of the electron-phonon interaction (EPI) according to which the spin susceptibility of metal is not affected by the EPI. According to this theory the figure of merit, so-called small adiabatic parameter, was estimated as $E_D/E_F \approx 10^{-2}$ [1 and references therein] where $E_D$ is the Debye energy, and $E_F$ is the Fermi energy. However, Kim showed that the impact of the EPI on the spin susceptibility of metals can be enhanced by two orders of magnitude if exchange interactions between electrons have been taken into account [1,2]. In other words, the effect of the EPI on magnetic properties of metallic systems, and *vice versa*, is much more significant than commonly believed. In fact, based either on calculations and/or on measurements several authors recently reported results that do not agree with the standard theory. In particular, M. S. Lucas et al. wrote [3]: "*The phonon densities of states of body-centered-cubic Fe-V alloys across the full composition range were studied by inelastic neutron scattering, nuclear resonant inelastic x-ray scattering, and ab initio calculations. Changes in the PDOS were revealed at crossing the Curie temperature*", B. Alling et al. based on disordered local moments molecular dynamics calculations concluded that [4]: "*Lattice vibrations strongly affect the distribution of local magnetic moment in paramagnetic Fe viz. they weaken their mean values*", and I. S. Tupitsyn et al. noticed that [5]: "*This theory neglects the effect of magnetism on lattice dynamics and fails to explain enhancement of the critical temperature in phonon-mediated superconductors.*"

From experimental viewpoint relevant candidates for testing the Kim's prediction are such magnetic systems in which their magnetism is greatly delocalized (itinerant). In order to select such candidates one can use the Rhodes-Wohlfarth criterion (plot). The criterion forecasts that a magnetic system is itinerant if the Rhodes-Wohlfarth ratio, RW=$q_c/q_s$, is greater than 1 [6]. Here $q_c$ is related to the effective paramagnetic moment, $p_{eff} = g\mu_B \sqrt{S(S+1)}$, by $q_c = S/2$, and $q_s$ to the magnetization in saturation, $M_s = q_s \mu_B$. $S$ stands for the effective spin number per atom, $\mu_B$ indicates the Bohr magneton and *g* is the Lange *g*-factor.

Sigma-phase (σ) Fe-based compounds e. g. Fe-Cr, Fe-V, Fe-Mo or Fe-Re, to name just binary ones having magnetic properties, are particularly well-suited to investigate the effect of magnetism on the lattice dynamics as, according to the Rhodes-Wohlfarth criterion, their magnetism is greatly itinerant because RW = ~4-13 [7,8]. Really, Mössbauer spectrometric (MS) studies on σ-FeX compounds (X=Cr,V) revealed that both spectroscopic parameters pertinent to the lattice dynamics, viz. the center shift as well as the recoil-free factor, displayed an anomalous behavior on crossing from the paramagnetic into magnetically ordered states [9,10]. Noteworthy, likewise an applied magnetic field seriously affected the lattice vibrations [10]. MS also revealed in the C14 Laves phase $Nb_{0.975}Fe_{2.025}$ a strong anomaly that starts at a transition from a paramagnetic to a ferromagnetic state [11]. Recently, using the Mössbauer effect on $^{119}$Sn nuclei diffused into metallic chromium (highly itinerant antiferromagnet) clear anomaly in the behavior of the center shift was detected at the Néel temperature [12]. In turn, MS measurements carried out on $^{57}$Fe nuclei in the temperature interval between 80 K and 350 K exhibited quite complex vibrational behavior of Fe atoms in the Cr matrix [13].

In this review are presented Mössbauer spectroscopic results that show, at least for itinerant magnets, that the lattice vibrations in a magnetic state are different than those in a paramagnetic state.

**2. Mösbauer spectroscopy and lattice dynamics**

Mössbauer spectroscopy (MS) has been successfully applied in investigations of lattice dynamics. However, as a resonant method, it gives information on the lattice vibrations as seen by resonant (probe) atoms. The most frequently used ones have been $^{57}$Fe and $^{119}$Sn isotopes. Two spectral parameters are relevant to the lattice dynamics viz. the center shift, *CS*, and the recoil-free fraction, *f*. In both cases their temperature dependence must be known in order to determine features pertinent to the lattice dynamics e. g. the Debye temperature, $T_D$.

**2.1. Center shift**

The temperature dependence of the center shift, *CS(T),* can be expressed by the following equation:

$$CS(T) = IS(0) + SOD(T) \qquad (1)$$

Where *IS(0)* stays for the isomer shift (temperature independent) and the second term in eq. (1), is known as the second-order Doppler shift, *SOD*. The latter is related to the mean-square velocity of the vibrating atoms, $<v^2>$, via the formula:

$$SOD = -\frac{E_\gamma}{2c^2}\langle v^2 \rangle \quad (2)$$

Where $E_\gamma$ stands for the energy of γ-rays (14.4 keV for $^{57}$Fe and 23.8 keV for $^{119}$Sn) and *c* is the light velocity.

In other words, the temperature dependence of *CS* goes via *SOD*. The latter, within the Debye model of the lattice vibrations, reads as follows:

$$SOD(T) = \frac{3k_B T}{2mc}\left(\frac{3T_D}{8T} - 3\left(\frac{T}{T_D}\right)^3 \int_0^{T_D/T} \frac{x^3}{e^x - 1}dx\right) \quad (3)$$

Where *m* is the mass of an $^{57}$Fe atom, $k_B$ is the Boltzmann constant, *c* is the speed of light, and $x=h\nu/k_B T$ (ν being the frequency of vibrations).

It follows from equations (1) and (2) that measurement of *CS(T)* gives information on the temperature dependence of the mean-square velocity of vibrations hence on the kinetic energy, $E_k$, of vibrations.

## 2.2. Recoil-free fraction, *f*

This quantity means the fraction of all Mössbauer gamma rays of the transition which are emitted or absorbed without recoil energy loss. It is related to the mean-square amplitude of vibrating atoms, $<x^2>$, via the following expression:

$$f = \exp[-(\frac{E_\gamma}{\hbar c})^2 \langle x^2 \rangle] \quad (4)$$

Assuming the lattice vibrations can be described by the Debye model, the recoil-free fraction can be related to the Debye temperature by the following formula:

$$f = \exp[-\frac{6E_R}{k_B T_D}\{\frac{1}{4} + (\frac{T}{T_D})^2 \int_0^{T_D/T} \frac{x}{e^x - 1}dx\}] \quad (5)$$

$E_R$ stands here for the recoil-free energy, $E_R = \frac{E_\gamma^2}{2mc^2}$.

Thus, the knowledge of $f$ enables determination of $<x^2>$, hence calculation of the potential energy of vibrations in the harmonic approximation, $E_p = \frac{1}{2} N <x^2>$, $N$ being the force constant.

## 3. Results

### 3.1. Sigma phase

The sigma ($\sigma$) phase (space group $D^{14}_{4h}$ - $P4_2/mnm$ ) is a member of a family of the Frank-Kasper (FK) phases [14]. The sigma-phase can be formed only in such alloys in which at least one constituting element is a transition metal. The unit cell of $\sigma$ is tetragonal and it accommodates 30 atoms which are distributed over five different lattice sites, usually denoted as A, B, C, D and E. The coordination numbers of the sites are high (12-16), and the distribution of atoms is not stoichiometric. These features, in combination with the fact that $\sigma$ can be formed in a certain range of composition, make it that $\sigma$-phase alloys show a diversity of physical properties that can be tailored by changing constituting elements and/or their relative concentration. Structural complexity and chemical disorder make them also an attractive yet challenging subject for investigation. The interest in $\sigma$ is further justified by a deteriorating effect on useful properties of technologically important materials in which it has precipitated [15,16]. It should be, however, added that attempts have been undertaken to profit from its high hardness and use profit from this feature to increase materials strength e. g. [17,18].

Concerning magnetic properties of $\sigma$ in binary alloys that can be studied by using MS, so far only $\sigma$ in Fe-Cr, Fe-V, Fe-Re and Fe-Mo alloy systems were revealed to be magnetic [7,19-21]. However, their magnetism has been recently shown to be more complex than initially anticipated (ferromagnetism). Namely, it has a re-entrant character with a spin-glass phase (SG) as the ground state [20-29]. Magnetic ordering (Curie) temperature, $T_C$, has values characteristic of a given alloy system, and for the given system on its composition. In particular, $T_C$ ranges between: ~8 K and ~39 K for $\sigma$-FeCr [7], ~10 K and ~335 K [19,24,29] for $\sigma$-FeV, ~20 K and ~42 K for $\sigma$-FeMo [21,22], ~21 K and ~53 K for $\sigma$-FeRe [26,28] and ~8 K and ~44 K for $\sigma$-FeCrNi [30]. It should be noted that the lower the value of $T_C$ the higher the value of the Rhodes-Wohlfart ratio, hence the greater the degree of delocatization of the magnetic moments. This in turn implicates, in the light of the Kim's theory, that the lower the $T_C$-value the stronger the effect of magnetism on the lattice dynamics.

### 3.1.1. Fe-Cr

The σ-phase in the Fe-Cr system is regarded as an archetype of σ. As can be seen on the relevant crystallographic phase diagram (Fig. 1), it can occur provided the Cr concentration lies between ~44 at.% and ~50 at.%. Two σ-$Fe_{100-x}Cr_x$ alloys were investigated viz. x=46 and 48 with respect to the effect of magnetism on the lattice dynamics [22]. Both the effect of the internal magnetism as well as that of an externally applied magnetic field was studied.

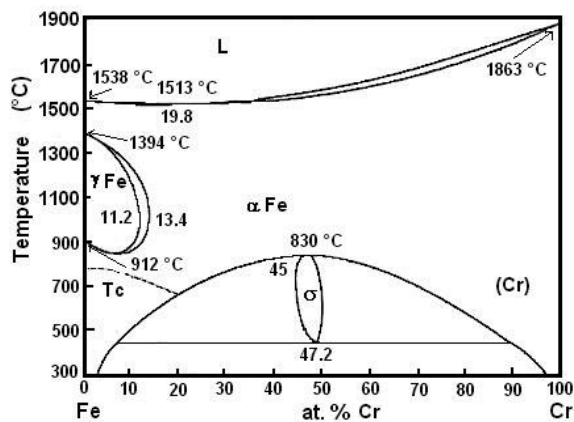

Fig. 1 Crystallographic phase diagram of the Fe-Cr alloy system after Baker H. Alloy phase diagrams handbook, Vol. 3, ASM International, 1992.

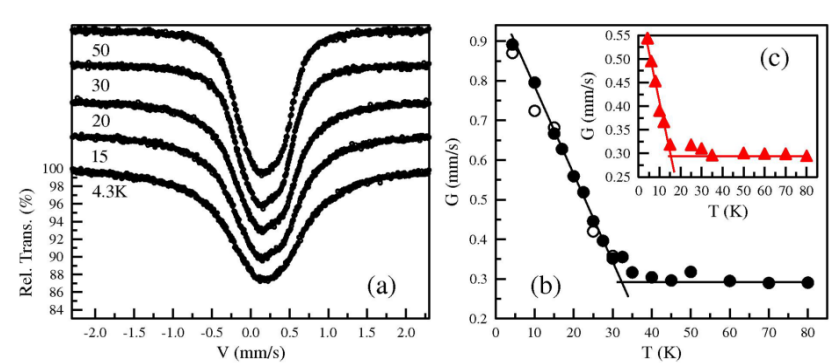

Fig. 2 (a) $^{57}$Fe Mössbauer spectra recorded on σ-$Fe_{54}Cr_{46}$ at various temperatures shown, and full line-width at half maximum, G, vs. temperature, T, for (b) σ-$Fe_{54}Cr_{46}$ and (c) σ-$Fe_{52}Cr_{48}$. The crossing between the horizontal line (paramagnetic phase) and the diagonal one (magnetic phase) indicates the Curie temperature, $T_C$, equal

to~33 K and ~15 K, respectively. In (b) open circles stand for the data obtained from repeated measurements [9].

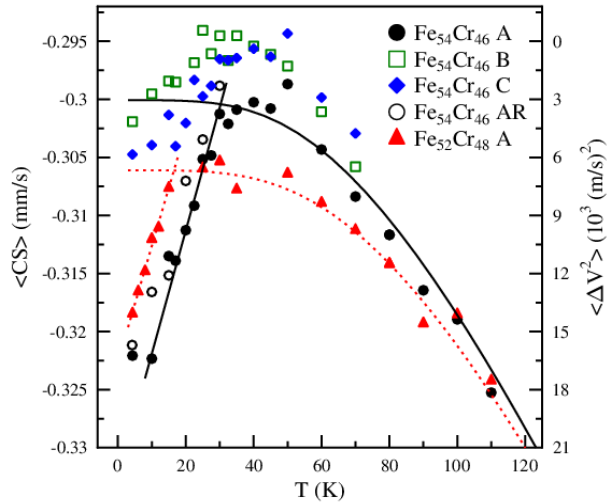

Fig. 3 Temperature dependence of the center shift, <CS>, as determined for two σ-$Fe_{100-x}Cr_x$ alloys with $x$=46 and 48 using different fitting procedures (A, B, C). AR stands for the data obtained by fitting with A the spectra obtained with a repeated measurements. Solid and dashed curves stand for the best-fit of the data obtained with method A to the Debye model [9].

Obviously, a significant drop from the Debye model prediction occurs for both alloys. The data within the drop shows a linear behavior and the temperature of the intersection of the lines with the Debye-like curves agrees well with the Curie temperature found for both samples. The right-hand axis is scaled with a change of the average square velocity, <$\Delta v^2$>, within the magnetic phase. Note that <$\Delta v^2$> increases on decreasing temperature.

An effect of an external magnetic field, $B_o$, was also investigated [9]. Namely, Mösbauer spectra were recorded at 4.2 K on both samples subjected to $B_o$ up to 13.5 T. The in-field spectra measured on σ-$Fe_{54}Cr_{46}$ can be seen in Fig. 4a.

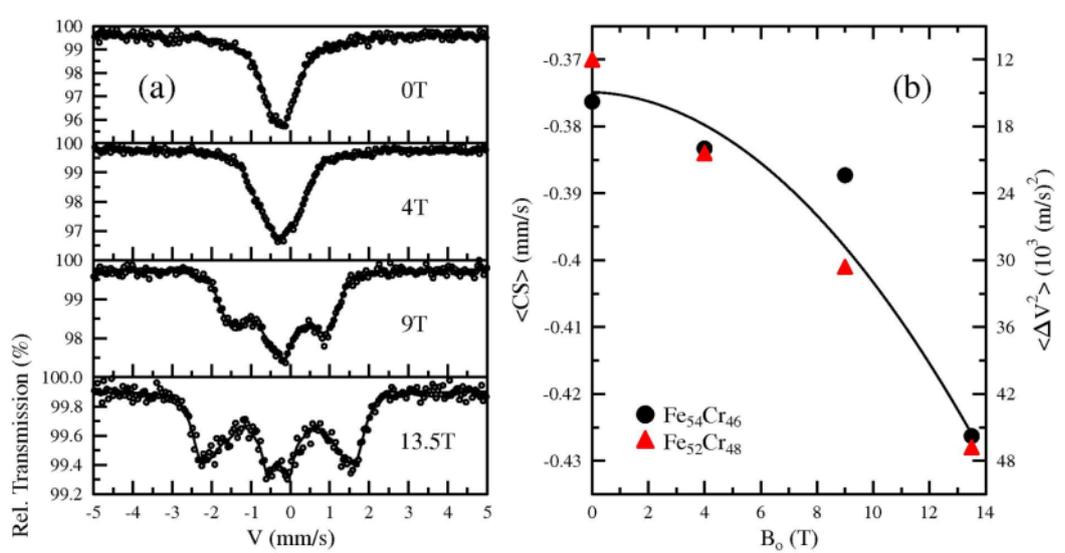

Fig. 4 (a) $^{57}$Fe Mössbauer spectra recorded at 4.2 K on σ-Fe$_{54}$Cr$_{46}$ in an external magnetic field of different values, as labelled. In (b) is displayed the dependence of the center shift, <CS>, on $B_o$ for both investigated samples. The solid line stands for the best parabolic fit to the data. The right-hand axis is scaled by a change of the average square velocity, <$\Delta v^2$> [9].

The effect of $B_o$ is presented in Fig. 4b. It is clear that its influence on <CS> is similar to that of the intrinsic magnetism i.e. the value of <CS> decreases with $B_o$ what signifies an increase of the kinetic energy of vibrating Fe atoms. One can therefore conclude that the increase of the kinetic energy of Fe atoms vibrations, $\Delta E_k$, can be caused both by the internal magnetism as well as by an external magnetic field. The maximum increase of $E_k$ for σ-Fe$_{54}$Cr$_{46}$ $\Delta E_k = E_k$ (4.2K)-$E_k$ (33K)≈4meV. The corresponding maximum decrease of the potential energy estimated from the change of the *f*-parameter is ~24 meV. Consequently, the total energy of Fe atoms vibrations decreased by $\Delta E \approx$ 20 meV on lowering temperature from ~33 K down to ~4 K. In other words, the magnetism makes vibrations of Fe atoms unharminic, suppresses vibrations amplitude but accelerates vibrations velocity.

Noteworthy, the related decrease of the thermal energy, $\Delta E_T = k_B \cdot \Delta T$ = 2.5 meV, hence 8 time less. This disparity between $\Delta E$ and $\Delta E_T$ clearly testifies to its magnetic origin.

### 3.1.2. Fe-V

The field of the σ-phase occurrence in the Fe-V alloy system in *x-T* coordinates is ~20 time larger than the one of σ in the Fe-Cr alloy – see Fig. 5. In particular at 600°C, the *x*-range of σ existence spans between ~32 at.% V and ~65 at.% V.

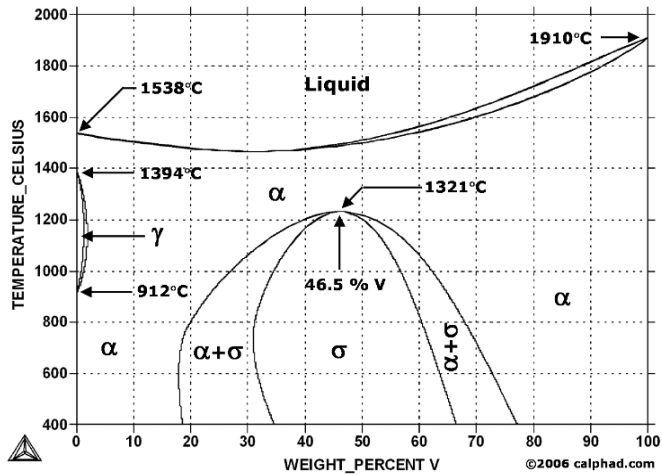

Fig. 5 Crystallographic phase diagram of the Fe-V alloy system [31].

The values of the Curie temperature of the σ-$Fe_{100-x}V_x$ alloys strongly depend on the V content e. g. $T_C$≈335 K for *x*=32 and $T_C$≈10 K for *x*=55 K – see Fig. 6.

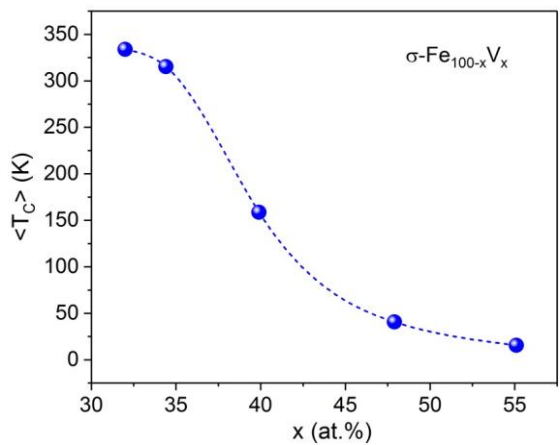

Fig. 6 Dependence of the average value of the Curie temperature, <$T_C$>, on the V content in σ-$Fe_{100-x}V_x$ alloys (average over the values obtained with magnetization and MS measurements). Plot made using the data published elsewhere [19,29,30].

Consequently, by changing the concentration of V one can efficiently tune the value of $T_C$. This, in turn, following the Rhods-Wolhfarth criterion, allows to ably control the degree of magnetic delocatization, and thus to see whether or not the effect the magnetism on the lattice dynamics depends on the degree of delocatization as follows from Kim's calculations [1,2].

Studies pertinent to the subject of the present paper were performed on samples with the following values of x=32, 34.4 and 39.9 i.e on the samples with rather quite high values of $T_C$, hence relatively low degree of magnetic delocatization. Consequently, the expected effect of magnetism on the lattice dynamics is not large.

### 3.1.2.1. Fe$_{68}$V$_{32}$

The recently constructed magnetic phase diagram of the σ-Fe$_{68}$V$_{32}$ alloy in the *H-T* plane (*H* – external magnetic field) gives a clear evidence on the re-entrant character of its magnetism with the record-high magnetic ordering temperature of 335(2) K [29].

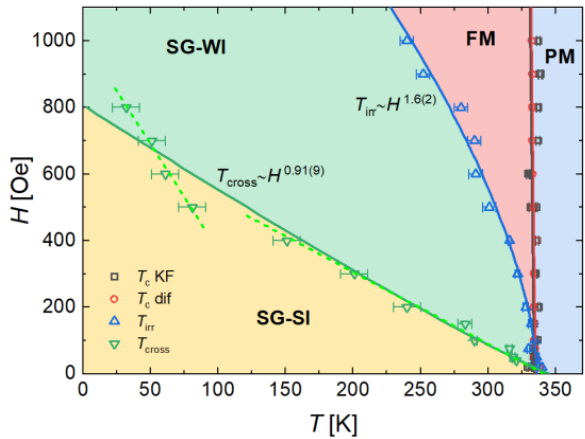

Fig. 7 Magnetic phase diagram of σ-Fe$_{68}$V$_{32}$ in the *H-T* plane. The solid and dashed lines represent the best fit to the corresponding data. PM stands for the paramagnetic phase, FM for the ferromagnetic phase, SG-WI and SG-SI are the spin-glass states with a weak and with a strong irreversibility, respectively [29].

$^{57}$Fe-site Mössbauer spectra were recorded on the σ-Fe$_{68}$V$_{32}$ sample in the temperature range of 5-390 K [31]. Examples of them can be seen in Fig. 8. Analysis of the spectra in terms of five components associated with five different lattice sites in the unit cell of σ permitted to determine temperature dependence of

the center shift for each component, $CS_k$, $k$=A, B, C, D, E. The resulting plot is displayed in Fig. 9.

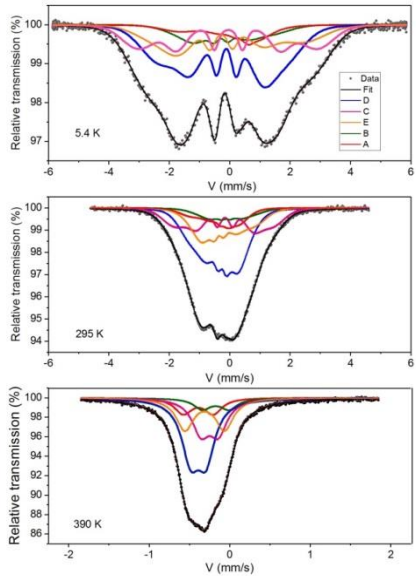

Fig. 8 $^{57}$Fe-site Mössbauer spectra recorded at different temperatures, as labelled, on the sample of σ-Fe$_{68}$V$_{32}$ [31]. The five subspectra corresponding to the A, B, C, D and E lattice sites are indicated, too.

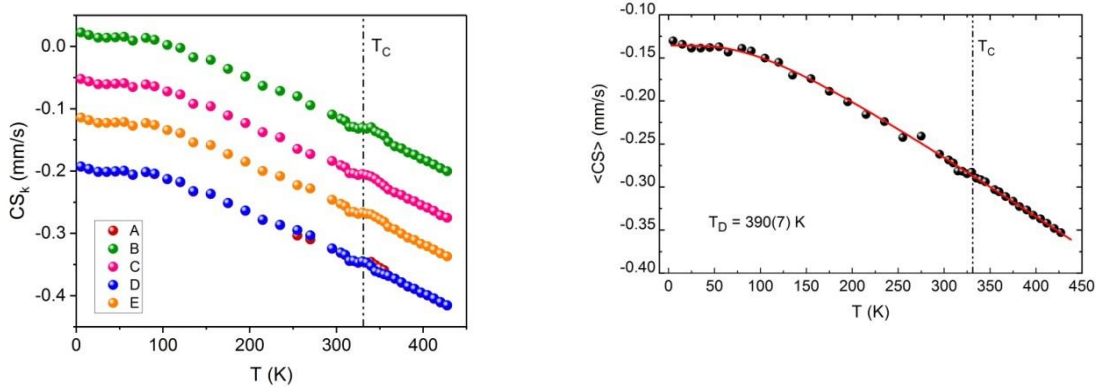

Fig. 9 Temperature dependence of: (left panel) $CS_k$ for the five spectral components $k$= A, B, C, D and E, and (right panel) the average center shift, <CS>. The horizontal dashed lines indicate the value of the Curie temperature, $T_C$ [31].

Some hardly seen disturbance at $T_C$ in the value of $CS_k$ can be seen for each component. Also some irregularity in the behavior is visible in each component for $T \lesssim$ ~50 K hence in the field of a strong-irreversibility of the spin-glass phase of the studied sample [29].

The results shown in Fig. 9 agree qualitatively with the Kim's prediction as, based on the Rhodes-Wohlfarth plot, the degree of itineracy of magnetism in the studied sample is very low. Also the so-called degree of frustration, $FD = \theta/T_C$, where $\theta$ is a paramagnetic Curie temperature, which is regarded as a proper figure of merit for determining the degree of itineracy, equals to 1.04 for the present case. Hence, it is very small.

### 3.1.2.1. Fe$_{60}$V$_{40}$

According to magnetization and Mössbauer-effect measurements, the Curie temperature for this compound is ~170 K [10,25], hence it is significantly less than the one for σ-Fe$_{68}$V$_{32}$, as discussed above. Consequently, the magnetism of σ-Fe$_{60}$V$_{40}$ should be much more itinerant. Indeed, the RW-ratio was estimated as high as 5.5, but the degree of frustration *FD*=1.05 [25], is not greater than the one found for σ-Fe$_{68}$V$_{32}$ [29]. The magnetism has a re-entrant behavior with the spin-freezing temperature of ~164 K [25]. This means that at room temperature the FM phase exists only in a narrow (*ΔT*~6 K) temperature range.

Regarding the Mössbauer-effect measurements, they were performed in the temperature range of 5-295 K [10]. Examples of the recorded spectra are shown in Fig. 9.

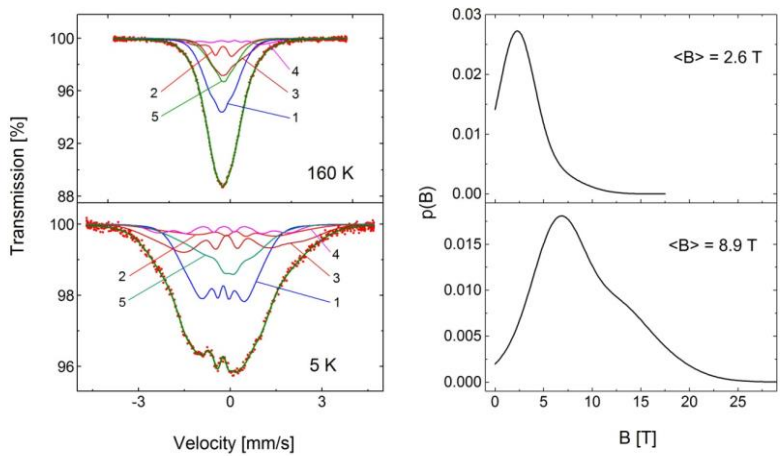

Fig. 9 Left panel: $^{57}$Fe Mössbauer spectra recorded on $\sigma$-Fe$_{60}$V$_{40}$ at 160 K and at 5 K. The five components corresponding to the five lattice sites are indicated. Right panel: Hyperfine field distributions curves derived from the spectra shown in the left panel. The average values of the hyperfine field, <B>, are displayed [10].

The analysis of the spectra yielded both spectral parameters relevant to the lattice dynamics viz. the average center shift, <CS>, and the *f*-factor.

### 3.1.2.1A. Center shift

Temperature dependence of <CS> is presented in Fig. 10. A discontinuity at $T \approx 168$ K, hence close to the magnetic ordering temperature, can be clearly seen. Consequently, the <CS> - data were fitted to eq. (1) separately for the paramagnetic and the magnetic ranges. Values of the Debye temperature derived therefrom are significantly different viz. 485(15) K for the paramagnetic phase and 322(17) K for the magnetic phase. This gives a clear evidence that the lattice vibrations seen through the probe Fe atoms are affected by the magnetism of the studied compound. It is worth noting that the vibrations in the whole temperature range are harmonic, at least as the kinetic energy of vibrations is concerned. Yet the smaller value of $T_D$ in the magnetic phase indicates a lattice softening in comparison to the paramagnetic phase.

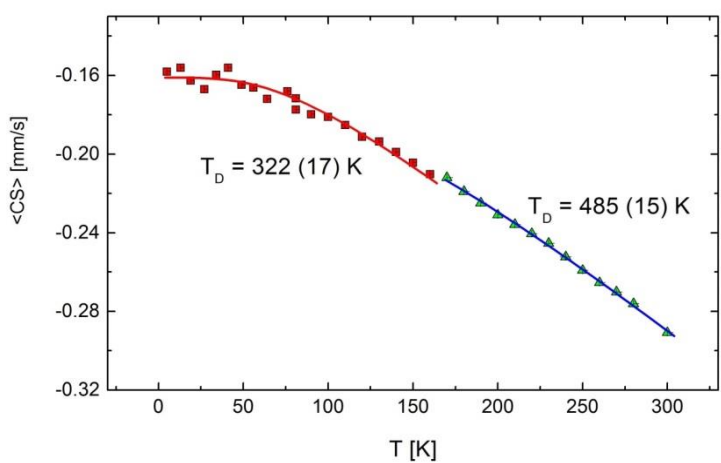

Fig. 10 Temperature dependence of the average center shift, <CS>. The best-fit curves in terms of eq. (1) are indicated for the paramagnetic and magnetic phases.

Values of the Debye temperatures obtained for the paramagnetic and magnetic phases are displayed [10].

### 3.1.2.1B. Recoil-free fraction

Determination of the *f*-factor, is technically much more difficult than that of the center shift. In a thin absorber approximation, the *f*-factor is proportional to a spectral area, *A*. In practice one uses a normalized spectral area, $A/A_o$, as a measure of the relative *f*-factor, $f/f_o$ ($A_o$ being the spectral area of the spectrum measured at the lowest temperature). One usually plots a temperature dependence of $ln(f/f_o)$. The plot obtained by authors of Ref. 10 can be seen in Fig. 11. The behavior is clearly very anomalous. On decreasing temperature from 295 K to ~180 K, hence in the paramagnetic phase, the dependence is pretty linear i.e. the lattice dynamics is harmonic. A linear behavior is also observed in a temperature range between ~180 K and ~150 K, yet with much larger slope. This temperature interval can be regarded as a phase with a ferromagnetic ordering. Below ~150 K, the $ln(f/f_o)$ dependence is not linear and two subranges denoted as I and II were distinguished.

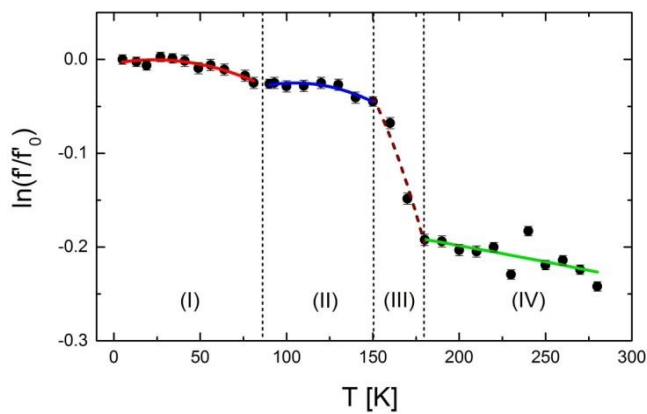

Fig. 11 Ln($f/f_o$) vs. temperature, *T*, for σ-$Fe_{60}V_{40}$ alloy. Five ranges, separated by dashed lines, can be identified: I and II in which the behavior is not linear, III and IV in which the behavior is linear. The lines stand for the best-fits to the data in particular ranges [10].

The plot shown in Fig, 11 reflects a re-entrant character of magnetism revealed in this compound [25]. Namely, I and II cover the temperature range in which a spin-glass occurs, III the one with a ferromagnetic ordering and IV coincides with the paramagnetic phase. The non-linearity seen in the spin-glass phase means that the vibrations of Fe atoms are anharmonic in the SG-phase. Their effect on $f$ can be expressed by the following equation [32,33]:

$$\ln f = -\frac{6E_R T}{T_D^2}(1+\varepsilon T+...) \quad (6)$$

Where $\varepsilon$ is an anharmonic coefficient.

Analysis of the data shown in Fig. 11 in terms of eq. (6) yielded for $T_D$ the following values: 657(150) K, 248(30) K, 104(20) K, 577(43) K for the ranges I, II, III and IV, respectively. Remarkably, the average value of $T_D$ over I, II and III (magnetic phase) is equal to 343(31) K which is similar to the value found from the temperature dependence of $<CS>$ in the magnetic phase.

The anharmonic coefficient, $\varepsilon = -2.3 \cdot 10^{-2}$ K$^{-1}$ for the range I and $\varepsilon = -4.6 \cdot 10^{-3}$ K$^{-1}$ for the range II. Remarkably, these values of $\varepsilon$ are very high. In particular, the former is 10-fold larger than the one determined for Fe impurities in the Cr matrix [13].

### 3.1.3. FeCrNi

Regarding ternary σ-phase alloys the only available results relevant to the subject of this paper depict an Fe$_{0.525}$Cr$_{0.455}$Ni$_{0.020}$ alloy [30]. The magnetic ordering temperature for this alloy equals to ~44 K [34], so according to the Rhodes-Wohlfarth criterion its magnetism should be greatly itinerant. In other words, one should expect a strong effect of the magnetism of the alloy on its lattice dynamics. Examples of the spectra recorded below and above $T_C$ are presented in Fig. 12.

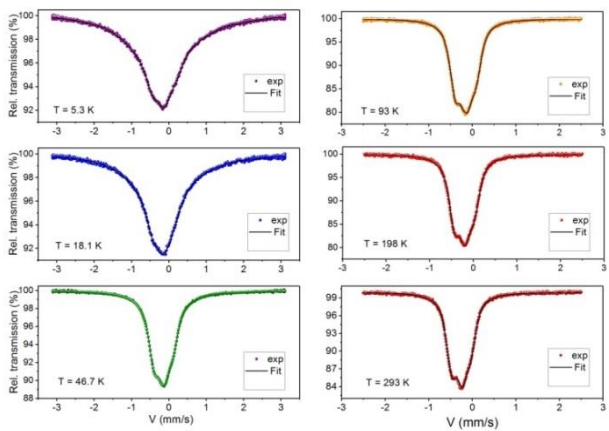

Fig. 12 A set of $^{57}$Fe Mössbauer spectra recorded on $Fe_{0.525}Cr_{0.455}Ni_{0.020}$ at various temperatures shown. Slight broadening seen at *T*=5.3 K and at 18.1 K reflects a weak magnetism of the sample [30].

Analysis of the spectra in terms of the hyperfine field distribution yielded values of the average hyperfine field, *<B>*, as well as that of the center shift, *<CS>*. The dependence of *<B>* on temperature is shown in Fig. 13, whereas that of *<CS>* in Fig. 14.

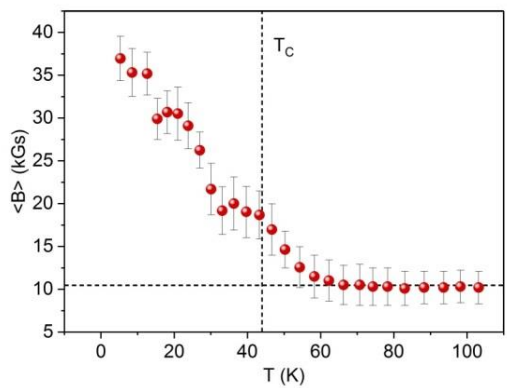

Fig. 13 The temperature dependence of the average magnetic hyperfine field, *<B>*. The vertical line marks the Curie temperature as determined by magnetization measurements [30].

It follows from Fig. 13 that an increase of *<B>* starts at temperature ~10 K higher than $T_C$. This effect can be explained in terms of differences both in the time window and a special resolution between the Mössbauer spectroscopy and magnetization measurements.

The temperature dependence of the average center shift, *<CS>*, determined by the analysis of the spectra is illustrated in Fig. 14.

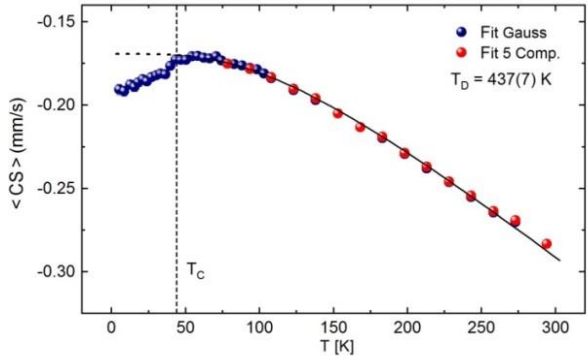

Fig. 14 The average center shift, <*CS*>, vs. temperature, *T*. The solid line represents the best fit of the data to eq. (3) (its extension into the magnetic phase is indicated by a dashed line). The value of the determined Debye temperature, $T_D$, is indicated. The vertical dashed line stands for $T_C = 43.8$ K, found with magnetization measurements [30].

A departure of the data from the trend predicted by the Debye model can be seen for $T \leq \sim 50$ K, hence in the magnetic phase of the studied alloy. It becomes greater with the decrease of *T*, hence with the increase of the magnetism strength. The decrease of <*CS*> (increase of its negative values) implies an increase of the average square-velocity of Fe atoms vibration in the lattice, $<v^2>$, hence an increase of the pertinent kinetic energy, $E_k$. From the thermodynamic viewpoint $E_k$ should be decreasing with decreasing *T*. The opposite behavior means that the increase of $E_k$ is at the expense of magnetism. The observed maximum increase of $<v^2>$ is equal to $\sim 2 \cdot 10^4$ (m/s)$^2$ and the related gain of $E_k$ amounts to ~6 meV. Noteworthy, the thermal decrease of $E_k$ due to the drop of temperature by ~45 K is 3.9 meV. This means that the net increase of $E_k$ on decreasing temperature from ~50 K to ~5 K amounts to ~10 meV.

### 3.2. C14 FeNb$_2$

The C14 hexagonal Nb$_{1-x}$Fe$_{2+x}$ compounds are known as Laves phase or the Frank-Kasper (ε, ρ or λ) phase. Its crystallographic phase diagram is displayed in Fig. 15.

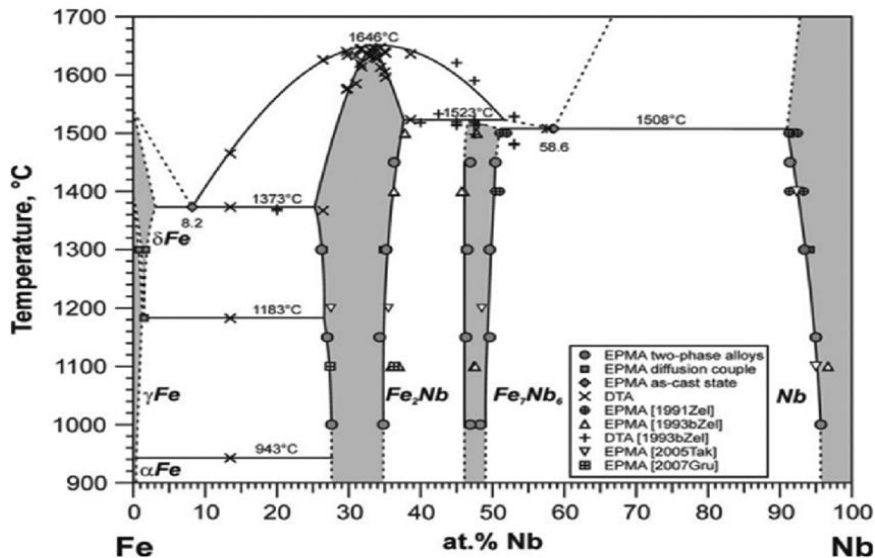

Fig. 15 Crystallographic phase diagram of the Fe-Nb alloy system [35].

A characteristic feature of NbFe$_2$ and other Frank-Kasper phases is a lack of stoichiometry. Instead they can be formed in a certain finite range of composition characteristic of a given phase. For example, the C14 ($\lambda$) Fe$_2$Nb compound can be formed with Nb content between ~35 at.% and ~46 at.%. This feature has a dramatic effect on magnetism of the phases, because by changing the composition one can effectively effect the strength (Curie temperature) of their magnetism.

To our best knowledge, there is only one available paper reporting on the effect of magnetism on the lattice dynamics in NbFe$_2$ [36]. Authors of this paper studied the Nb$_{0.975}$Fe$_{2.025}$ compound. Following magnetization measurements, the magnetism of this compound has a re-entrant character viz. on lowering temperature PM→FM→SG transition occurs. The compound orders ferromagnetically at ~63 K, is itinerant (RW=2.7) and magnetically weak [37]. In other words, it is a proper candidate for studying the effect of magnetism on the lattice dynamics.

The Mössbauer-effect measurements were carried out in a wide temperature range viz. 5-300 K [36]. Examples of the recorded spectra are presented in Fig. 16.

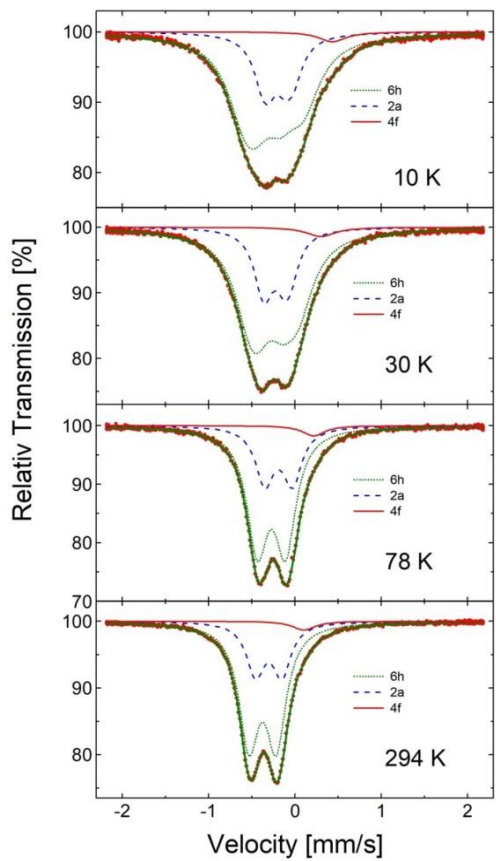

Fig. 16 Examples of the $^{57}$Fe Mössbauer spectra recorded at various temperatures shown. Three sub spectra associated with the 2a, 6h and 4f lattice sites populated by Fe atoms are indicated [36].

The spectra recorded at 10 K and 30 K show some broadening relative to the ones measured at 78 K and 294 K – see Fig. 17 (left). The broadening reflects a weak magnetism of the studied compound. Site-resolved values of the hyperfine field are illustrated in Fig. 17 (right). The maximum value of *B* at the site 6h is by a factor of two stronger than the value of *B* at site 2a.

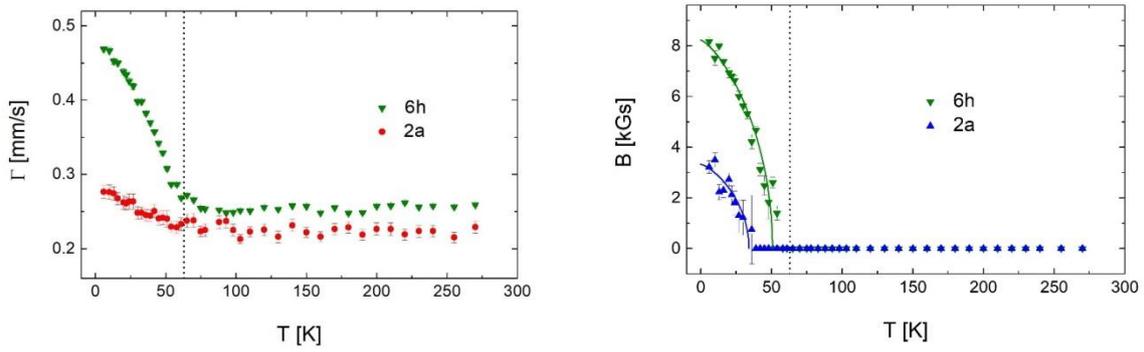

Fig. 17 (left) Line width at half maximum, Γ, and (right) related magnetic hyperfine field, *B*, vs. temperature, *T*, for the sites 2a and 6h. The vertical broken line stands for the Curie temperature found with the magnetization measurements [36].

Analysis of the spectra enabled determination of the center shifts for the two major components i.e. those related to the lattice sites 2a and 6h. Their temperature dependence is displayed in Fig. 15. As expected, the behavior is greatly anomalous for both sites. In both cases significant departure from the harmonic behavior takes place. The temperature at which the anomalies start to occur is close to the magnetic ordering temperature of the compound. Consequently, the anomalies can be associated with the effect of magnetism. Interestingly, the anomalous parts of *CS* is characteristic of the site. The data exhibiting a regular behavior (paramagnetic phase) were analyzed in terms of the Debye model – eq. (3), yielding the values of the Debye temperature viz. 544(10) K for 2a and 453(5) K for 6h. The different values of $T_D$ latter mean that although Fe atoms present on 2a and 6h sites vibrate harmonically in the paramagnetic phase, yet with different values of the spring constant.

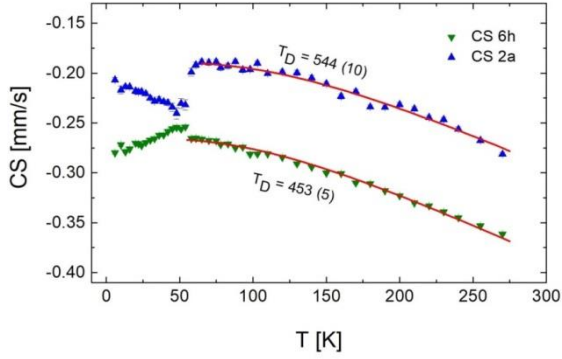

Fig. 18 Temperature dependence of the 2a and 6h sites center shift, *CS*. The best-fits to the regular parts of the data are indicated by solid lines labelled by the derived therefrom values of the Debye temperature, $T_D$ [36].

The effect of magnetism on the vibrations of Fe atoms can be also visualized by considering the temperature dependence of the average center shift, *<CS>*, as illustrated in Fig. 16.

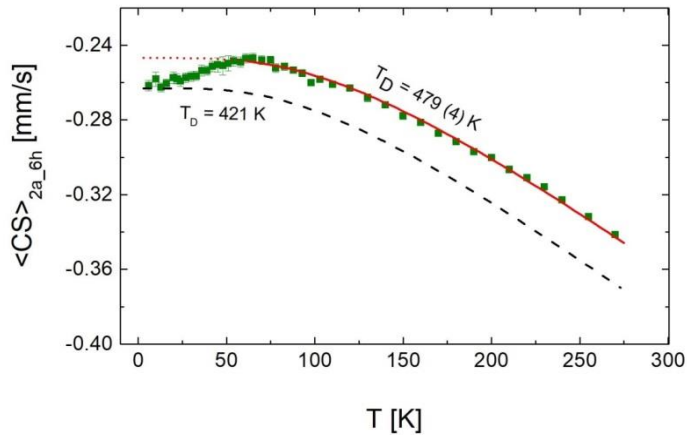

Fig. 16 Temperature dependence of the average (over 2a and 6h sites) center shift, $<CS>_{2a\_6h}$. The best-fit to the regular part of the data is marked by a solid line. The derived therefrom Debye temperature, $T_D$=479(4) K. The behavior of the average *CS* corresponding to $T_D$=421 K is marked by a dashed line [36].

Evidently, at $T\approx65$ K, hence close to the Curie temperature, the $<CS>_{2a\_6h}$ data significantly deviates from the trend seen at higher temperatures (paramagnetic phase). As discussed above, the drop of the center shift means an increase of the kinetic energy of vibrations. It also means a weakening of the lattice. Assuming the <CS>$_{2a\_6h}$ value at 5 K lies on the curve in line with the Debye model (dashed line in Fig. 16), the related value of $T_D$ would be 421 K.

### 3.3. Metallic chromium

Metallic chromium is regarded as an archetype of an antiferromagnet system (AF). Its Néel temperature $T_N \approx 313$ K [38]. Yet the ordering of magnetic moments is not characteristic of a normal AF because the moments, μ, have no constant magnitude. Instead they are harmonically modulated in space i.e. $\mu = \mu_1 sin\alpha + \mu_3 sin3\alpha$, where α=**q**·**r**, **q** is a wave vector and **r** is a distance in the real space, $\mu_1$ is an amplitude of the first-order harmonic and $\mu_3$ the one of the third-order. The $\mu_3/\mu_1$ ratio lies between 1.5 and 2.5 % depending on temperature [39,40]. The harmonic modulation of $\mu$ in the real space gives rise to name the effect as spin-density waves (SDWs). The importance of the SDWs follows from their relationship to the topology of the Fermi surface [40]. There are two types of the SDWs viz. (a) longitudinally polarized, LSDWs, with **q** ∥ μ, and (b) transversely polarized, TSDWs, with **q** ⊥ μ. The former can be called a low-temperature phase of AF, as it exist below the so-called spin-flip temperature, $T_{SF}$=123 K [40], and for the latter the term of a high-temperature AF phase is appropriate, as it occurs between $T_{SF}$ and $T_N$. The theory of Kim does not include AF ordering, but AF of chromium is highly itinerant, so it is of interest to see whether or not lattice vibrations are affected by the itinerant AF. Concerning the application of MS toward this end, one has to insert probe atoms into the Cr matrix in order to record Mössbauer spectra. The most popular Mössbauer isotopes are $^{57}$Fe and $^{119}$Sn, and so far, to our best knowledge, only they were used in the studies of chromium via MS (Ref. 12,40-45 for $^{119}$Sn and Ref. 13 for $^{57}$Fe). The lattice dynamics of Cr was studied both by recording the $^{119}$Sn spectra in the temperature interval few degrees above and below the Néel temperature [12] as well as the $^{57}$Fe spectra in the temperature range of 80-350 K [13].

### 3.3.1. $^{119}$Sn study

A set of the spectra recorded at various $T$ in the vicinity of $T_N$ on a single-crystal sample of Cr diffusion-doped with ~0.2 at.% $^{119}$Sn isotope is presented in Fig. 17 together with the hyperfine magnetic field (spin density) distribution histograms, p(H), derived therefrom. An appearance of a magnetic component can be seen on lowering T both in the spectra viz. as a broadening, and in the histograms in form of an additional peak (blue in the on-line version) characteristic of the SDWs.

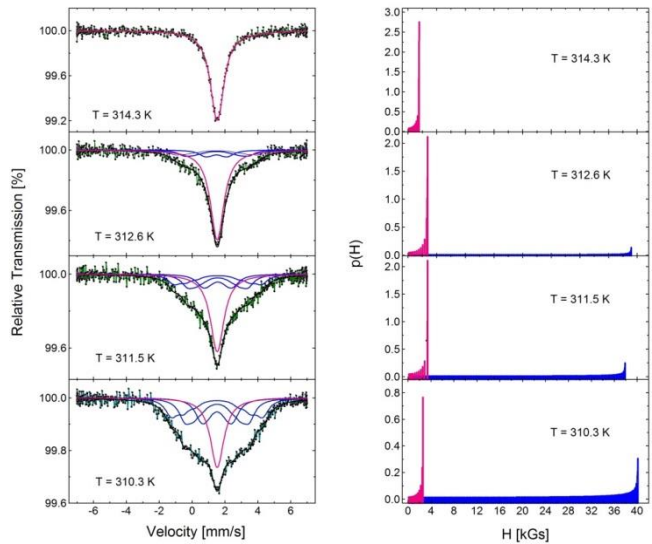

Fig. 17 (left panel) Examples of $^{119}$Sn spectra recorded at various temperatures on the sample of a single-crystal Cr sample doped with ~0.2 at.% $^{119}$Sn. Subspectra involved in their analysis with method A are indicated; (right panel) Histograms of the hyperfine field distribution derived from the corresponding spectra using method B. They reflect the increase of the magnetic component (in blue in the on-line version) on the decrease of T [12].

The spectra were analyzed with two different fitting procedures called A and B. Their detailed description can be found elsewhere [12]. Both ways of the spectra analysis yielded temperature dependence of the center shift, CS(T), a parameter of merit relevant to the lattice vibrations and related to the kinetic energy of vibrations, $E_k$. The CS(T) dependence obtained based on the fitting procedure A is shown in Fig. 18a while the one retrieved using the procedure B can be seen in Fig. 18b.

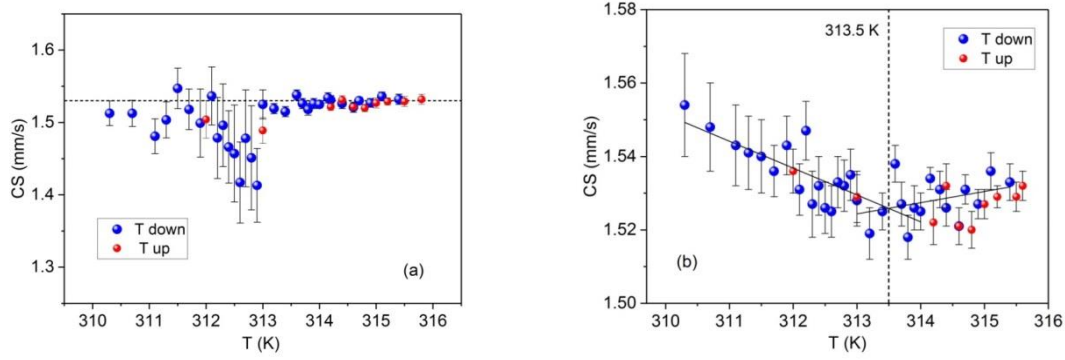

Fig. 18 Temperature dependence of the center shift, *CS*, as found with: (a) procedure A, and (b) procedure B. The horizontal dashed line in (a) stays for the average *CS*-value in the paramagnetic state. The anomalous behavior concerns *CS*-values of the magnetic component. Solid lines in (b) stay for the best linear fit to the data in the paramagnetic and in the antiferromagnetic state, respectively. The vertical dashed line indicates the crossing of the two straight lines [12].

Evidently, the data shown in Figs. 18a and 18b give evidence that there is an anomaly in *CS* at $T \approx 313$ K. The latter agrees well with the Néel temperature of chromium, hence it is justified to associate the anomaly with the PM→AFM transition.

Temperature changes of *CS* can be used to determine underlying changes of the kinetic energy, $\Delta E_k = 0.5 \cdot m \cdot \Delta <v^2>$, where m is the mass of the vibrating atom (here $^{119}$Sn) and $\Delta <v^2>$ is a change of the average square velocity of vibrations. The latter can be determined as $\Delta <v^2> = -2c \cdot \Delta SOD$, where $\Delta SOD$ stands for the corresponding change of the second-order Doppler shift.

Using the above given equations the change of *CS(T)* in the temperature range of 310.3 – 313.5 K, hence the antiferromagnetic state, was calculated as equivalent to a decrease of the kinetic energy of lattice vibrations by 8.5 meV, as detected by the probe $^{119}$Sn atoms. On the other hand, the change of *CS(T)* in the temperature interval of 313.5-315.6 K i.e. in the paramagnetic state, is equivalent to an increase of the kinetic energy of lattice vibrations by 2.2 meV. Thus, the kinetic energy of vibrations in the paramagnetic state close to the Néel temperature increases with the decrease of *T* at the rate of ~1meV/K, whereas this energy decreases in the antiferromagnetic state close to the Néel temperature at the rate of ~2.7 meV/K i.e.

much faster. Noteworthy, the corresponding changes of the vibrational energy due to the decrease of temperature, $\Delta E = k_B \cdot \Delta T$, are -0.27 meV and -0.18 meV, respectively.

The above-given results make evident that the lattice vibrations in metallic chromium are significantly affected by the magnetism of the system. One may argue that the present experiment does not give a direct evidence on the lattice vibrations of Cr atoms themselves because the measurements were carried out on Sn atoms introduced as impurities into the Cr matrix. Hence the vibrations of Sn atoms, which are much heavier than the matrix Cr atoms, may not properly reflect the vibrations of the Cr matrix itself. However, based on our previous studies on the magnetism of chromium using the $^{119}$Sn Mössbauer spectroscopy we found that the Mössbauer results are in line with those revealed with other methods on samples of pure Cr. In particular, (1) the value of the amplitude and the sign of the third-order harmonics of the SDWs [39] agree with the corresponding values revealed by the neutron diffraction (ND) study, (2) the value of the Néel temperature agrees within ±1 K with the value found from the ND study on a metallic chromium [38], (3) the value of the spin-flip temperature [45] agrees well with that determined with ND, and (4) the anomaly in *CS(T)* reported in [12] takes place at temperature that coincides with $T_N$ of a pure Cr. Based on all these findings one can quite confidently conclude that the vibrations of Sn atoms embedded into the Cr matrix correctly reflect the vibrations of the Cr matrix. In other words, the anomaly in *CS(T)* shown in Fig. 2 reflects the disturbance in the lattice vibrations of the matrix Cr atoms that can be associated with the transition from the paramagnetic into the antiferromagnetic state of the investigated system.

### 3.2.2. $^{57}$Fe study

$^{57}$Fe were performed on a metallic sample of Cr doped with 0.1 at.% Fe enriched to 95% in the $^{57}$Fe isotope [13]. It should be noted that Fe atoms, contrary to Sn ones, are magnetic $\mu_{Fe}$. It is known that even single Fe atoms embedded into the Cr matrix have quite substantial magnetic moments viz. $\mu_{Fe}$=1.4 $\mu_B$ [46]. Consequently, they behave differently than non-magnetic Sn atoms. The difference can be readily seen in the shape of the Mössbauer spectrum: while the one recorded at $^{119}$Sn is magnetically split by the nuclear Zeeman effect and its shape is characteristic of the

incommensurate SDWs [39], the spectrum recorded at $^{57}$Fe has a form of a slightly broadened singlet – see Fig. 19. A possible reason for the latter is a compensation of spins between the one of Fe and that of the SDWs.

In order to study a possible influence of AFM of Cr on the lattice dynamics of Fe atoms, the spectra, some of which can be seen in Fig. 19, were recorded in the temperature range between 80 K and 350 K. Based on the analysis of the spectra the relative recoil-free factor, $f/f_o$, was determined, a quantity related to the relative average square amplitude of vibrations, $<\Delta x^2>$, by $f/f_o = \exp[-k^2 <\Delta x^2>]$, $k$ being the wave vector of the 14.4 keV gamma rays.

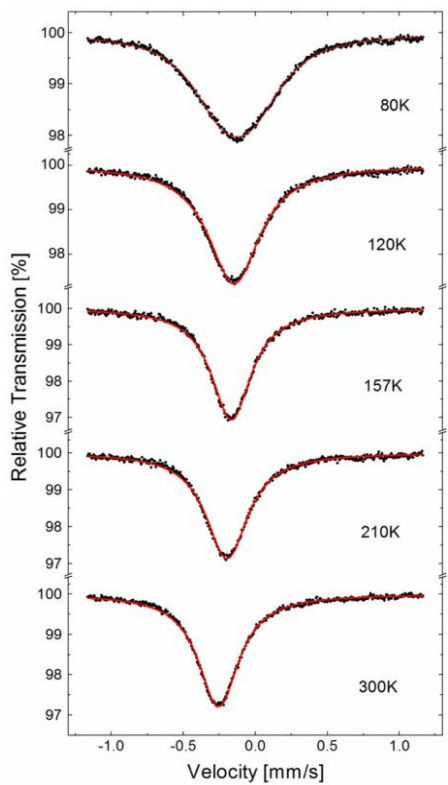

Fig. 19 A selection of $^{57}$Fe Mössbauer spectra recorded at different temperature as indicated. Note that all shown spectra were measured in the AFM phase of the sample [13].

The temperature dependence of $-\ln f/f_o$ is presented in Fig. 20.

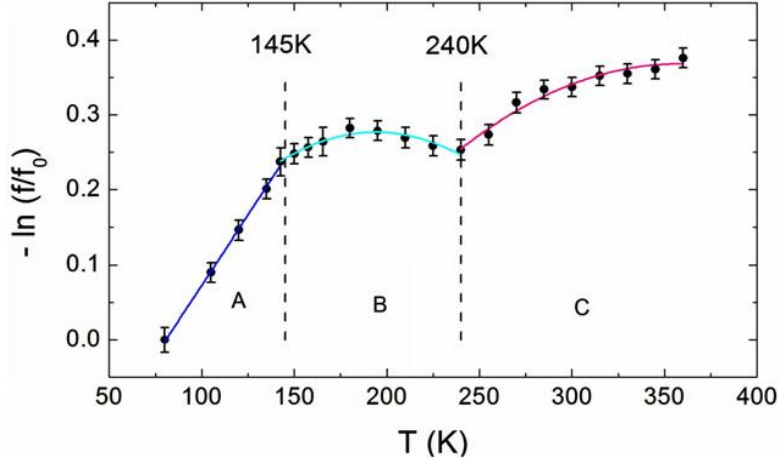

Fig. 20 *T*-dependence of -*lnf/f₀*. The solid lines represent the best-fit of the data to eq. (7) [13].

The *lnf/f₀ T*-dependence was analyzed in terms of the following equation:

$$ln\frac{f}{fo} = -lnf_o + (1 + \varepsilon T) \qquad (7)$$

One can readily notice that three ranges designated in Fig. 20 as A, B and C can be distinguished. In the low temperature range (A), the behavior is linear ($\varepsilon$=0) meaning that the vibrations of Fe atoms are harmonic. Noteworthy, this temperature interval quite reasonably coincides with the transversally polarized SDWs. The derived value of $T_D$=190(3) K. The ranges B and C cover the temperature interval in which the polarization of the SDWs is longitudinal in a pure Cr [40]. The data shown in Fig. 20 give evidence that the *lnf/f₀* dependence both in B as well as in C is nonlinear ($\varepsilon \neq 0$). The derived values of the Debye temperature are 155(2) K and 152(3) K, for B and C, respectively. The obtained values of $T_D$ are remarkably low in comparison with the $T_D$-value for pure Cr viz. 606 K at 0 K [48] and 424 K at 295 K [49]. They are also significantly smaller than the value of $T_D$=454(50) K determined based on the temperature dependence of <CS> measured in the range of 80-300 K using $^{119}$Sn as probe atoms in Cr [50]. The difference in the values of the Debye temperature of Cr found using as probe $^{119}$Sn and $^{57}$Fe atoms gives evidence that the dynamics of the two kinds of atoms is distinctly different. The most likely reason for this is that Sn atoms have no own magnetic moments whereas Fe atoms do have, even as single impurities embedded into the Cr matrix [46]. Theoretical calculations predict that non-magnetic impurities do not disturb the SDWs in chromium, while the magnetic ones have a pinning effect [51-54]. Nevertheless, experimental results obtained using both types of the probe atoms i.e. $^{119}$Sn and $^{57}$Fe gave evidence that their dynamics in the Cr matrix is affected by the antiferromagnetism of the matrix.

## 4. Iron mono-arsenide, FeAs

FeAs can be formed in a very narrow compositional range and has the MnP-type crystal structure (space group *Pnam* or *Pna2$_1$*) [55]. FeAs orders antiferromagnetically at 70 K [56] and the AFM state is constituted by itinerant incommensurate SDWs [57]. Consequently, it can be regarded as a good candidate to study the effect of magnetism on the lattice dynamics. Indeed, Blachowski et al. carried out systematic Mössbauer-effect measurements in the temperature range of 4.2-300 K and, additionally, in a furnace at 600 K, 800 K and 1000 K [58].

In the analysis of the spectra two different positions of Fe atoms in the unit cell were taken into account. This permitted to obtain site-resolved values for the center shift and for the hyperfine field. Temperature dependence of the latter is visualized in Fig. 21.

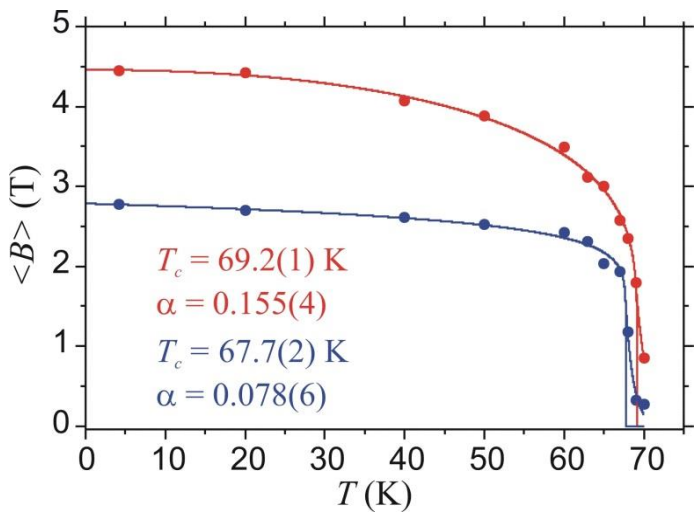

Fig. 21 The average hyperfine field, <*B*>, as obtained for FeAs using two-component model (for different Fe sites in the lattice) in the spectra analysis. $T_C$ stands for the magnetic ordering temperature (in fact the Néel temperature), and α denotes the static critical exponent [58].

It is noteworthy that the magnetic ordering temperature (Néel) is within the error limit the same for the two sites, yet the maximum values of the field are significantly different. Results depicting the center shift, indicated by *S,* are displayed in Fig. 22.

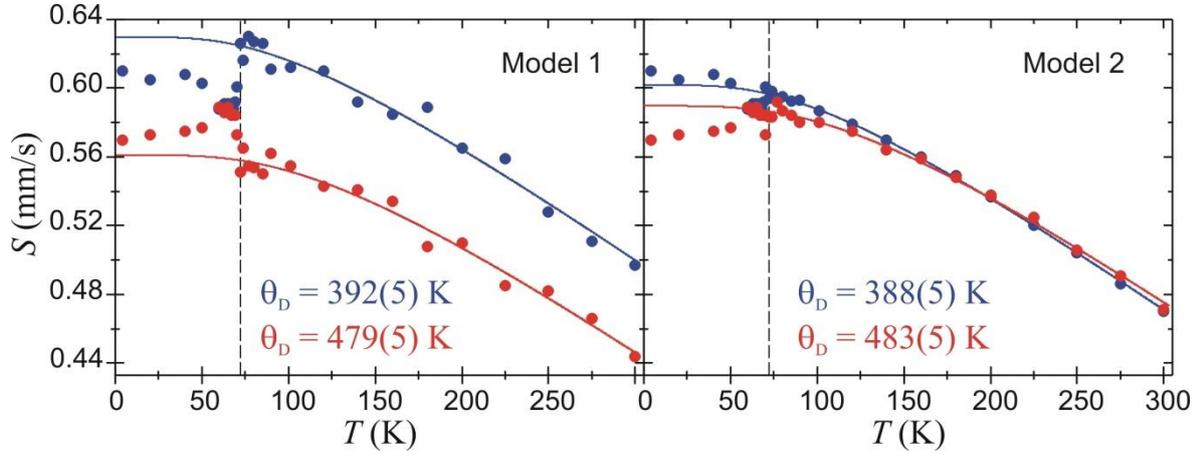

Fig. 22 Temperature dependence of the center shift, *S*, as found for the two spectral components with the model 1 (left) and model 2 (right) of the spectra analysis. The vertical line indicates the Néel temperature. For both sets of the data a significant deviation from the trend seen in the paramagnetic phase ($T \geq$ ~70 K) exists in the AMF phase ($T \leq$ ~ 70 K). Debye temperature values determined from the data in the PM phase are displayed, too [58].

There are two plots of the temperature dependence of *S*, because the spectra were fitted with two different models, model 1, and model 2 (for details see [58]). Both plots give evidence that below the Néel temperature (indicated by vertical line) a departure occurs from the trend seen in the paramagnetic phase ($T > T_N$). Alike behavior was also seen for $NbFe_2$ – Fig. 18. It signifies unharmonic vibrations of Fe atoms in the AFM phase. The data in the PM-phase were fitted to eq. (3) and the Debye temperature was obtained for both lattice sites. Its value, indicated in Fig. 22, is characteristic of the site and rather independent of the model applied in the spectra analysis.

Also a second spectral parameters pertinent to the lattice vibrations i.e. the recoil-free fraction was considered [58]. Its relative average value, denoted as *<f>/<f$_o$>*, is plotted vs. temperature in Fig. 23. A very sharp increase of *<f>/<f$_o$>* can be seen at ~70 K with a saturation-like shape for lower temperatures. The increase means a decrease of the average square amplitude of lattice vibrations as seen via the probe Fe atoms. The behavior of <f>/<f_o> in the paramagnetic phase is regular i.e. in line with the Debye model. The fitting of the data in the PM range to eq. (2) yielded value of the Debye temperature, $\theta_k$. Three different values of $\theta_k$ are displayed in Fig. 23 as a result of two different ways of fitting. Namely, $\theta_{iso}$=368(9) K for the harmonic isotropic model of vibrations, $\theta_{ac}$=330(1) K and $\theta_b$=508(68) K for the harmonic anisotropic model of vibrations.

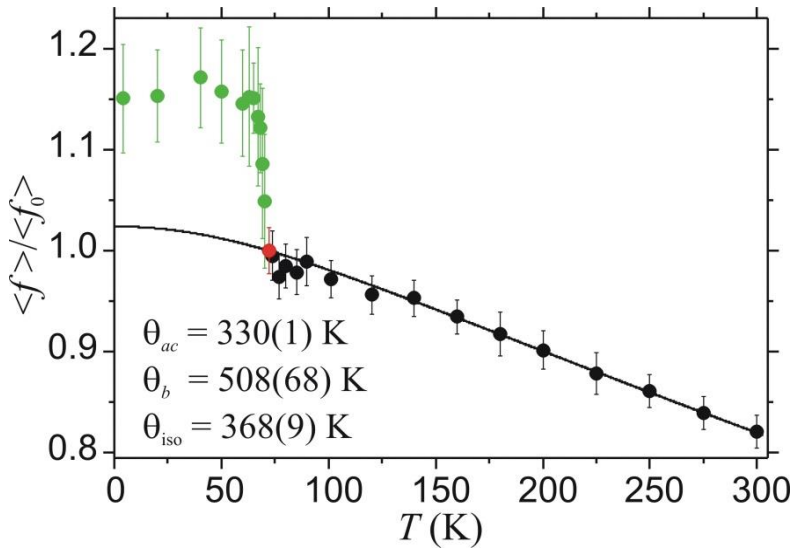

Fig. 23 The relative average recoil-free fraction, <f>/<f_o>, plotted vs. temperature, $T$. Note a sharp increase at $T \approx 70$ K indicative of the unharmonic vibrations of Fe atoms in the AFM phase. The solid line stands for the best-fit of the data in the PM state to eq. (2). Various values of the Debye temperature, $\theta_k$, are shown. Their meaning is explained in the text [58].

## 5. Summary and Conclusions

Accessible Mössbauer spectroscopic results relevant to the effect of magnetism on lattice dynamics have been summarized. In particular, the following metallic systems are presented and discussed: (1) $\sigma$-$Fe_{100-x}Cr_x$ ($x$=46 and 48), (2) $\sigma$-$Fe_{100-x}V_x$ ($x$=32 and 40), (3) $\sigma$-$Fe_{0.525}Cr_{0.455}Ni_{0.020}$, (4) $\lambda$-$Nb_{0.975}Fe_{2.025}$, (5) metallic Cr and (6) Iron mono-arsenide, FeAs. For all cases Mössbauer-effect measurements were performed at $^{57}$Fe nuclei, and, in addition, at $^{119}$Sn nuclei for (5). Center shift, <CS>, and the recoil-free fraction, $f$, were considered as parameters of merit. The former is related to the average square velocity, hence the kinetic energy, of vibrations and the latter to the average square amplitude of vibrations, hence the potential energy of vibrations in the harmonic approximation. The common feature of the presented systems is an itinerant character of their magnetism. In all cases anomalies in the temperature dependence of <CS> and/or *f* occur. The size of the anomalies for the Frank–Kasper phases (σ and λ) is correlated with the value of their Curie temperature, $T_C$: the lower $T_C$ the bigger the anomaly. Such behavior is in line with the Rhodes-Wohlfarth criterion of the itineracy of magnetism and the prediction of the Kim's theory. In the magnetic phase, the lattice vibrations, as seen by $^{57}$Fe and $^{119}$Sn probe atoms, are in generally unharmonic. In addition, values of the Debye temperatures, $T_D$, are smaller than those in the paramagnetic phase of a given alloy. This feature indicates a lattice softening in the magnetic state. Noteworthy, the $^{57}$Fe and $^{119}$Sn results obtained for chromium are significantly different. In particular, $T_D$=454(50) K as calculated based on the *CS(T)* determined from the $^{119}$Sn Mössbauer spectra, and <$T_D$>=166 K (average over three different values) as found from the *f*(T)-dependence derived from the $^{57}$Fe spectra. The case of chromium is, however, exceptional because its magnetism is constituted by spin-density waves, SDWs, related to a topology of the Fermi surface. Their properties can be substantially distorted by magnetic impurities like Fe atoms. This is most likely the reason why the Cr lattice dynamics revealed by the experiment on $^{119}$Sn atoms is not the same as that found by the measurements performed on $^{57}$Fe atoms. The magnetism of the iron mono-arsenide, FeAs, is similar to that of Cr i.e. it is constituted by incommensurate spin-density-waves. For FeAs clear-cut indications of unharmonic vibrations of Fe atoms have been revealed both in the center shift as well as in the recoil-free fraction.

In summary, the Mössbauer-effect measurements reported in this paper give a clear-cut evidence that both internal magnetism as well as externally applied magnetic field have a noticeable impact on lattice vibrations.

223902